\documentclass[letter,twocolumn]{jpsj2}

\title{Gauge Field Formulation of Adiabatic Spin Torques}

\author{Hiroshi \textsc{Kohno}$^{1}$
\thanks{E-mail address: kohno@mp.es.osaka-u.ac.jp} 
and Junya \textsc{Shibata}$^{2}$
\thanks{E-mail address: jshibata@postman.riken.go.jp}
}

\inst{$^{1}$Graduate School of Engineering Science, Osaka University,
Toyonaka, Osaka 560-8531, Japan \\
$^{2}$RIKEN-FRS, 2-1 Hirosawa, Wako, Saitama 351-0198, Japan 
}

\recdate{\today}

\abst{Previous calculation of spin torques for 
small-amplitude magnetization dynamics 
around a uniformly magnetized state 
[J. Phys. Soc. Jpn. {\bf 75} (2006) 113706] 
is extended here to the case of finite-amplitude dynamics. 
 This is achieved by introducing an \lq\lq adiabatic'' spin frame 
for conduction electrons, 
and the associated SU(2) gauge field. 
 In particular, the Gilbert damping is shown to arise from the 
time variation of the spin-relaxation source terms in this new 
frame, giving a new physical picture of the damping. 
 The present method will allow a \lq\lq first-principle'' derivation 
of spin torques without any assumptions such as rotational symmetry 
in spin space.}

\kword{Landau-Lifshitz-Gilbert equation, $s$-$d$ exchange interaction, 
spin torque, Gilbert damping, spin current, spin relaxation, 
magnetic impurity, local spin frame, gauge field}

\begin{document}
\maketitle

 Magnetization dynamics driven by electric/spin current 
in submicron-scale ferromagnets has been a subject of active study 
for a decade. 
 They include magnetization reversal in multilayer systems 
\cite{Slonczewski96,Berger96,multilayer_review}, 
displacive and resonant motions of domain walls in thin wires 
\cite{Berger84&92,
Yamaguchi04,Yamanouchi04&06,Saitoh04,Klaui05,Hayashi06&07,Thomas06,
TK04,Zhang05,Thiaville05,Barnes05,TTKSNF06}, 
and those of magnetic vortices realized in disks 
\cite{Shibata06,Ishida06,Kasai06,Yamada07}. 
 These phenomena are understood as caused by spin torques, 
the effect that conduction electrons exert on magnetization 
through the microscopic $s$-$d$ exchange interaction.

 There are several kinds of spin torques. 
 For slow and smooth dynamics of magnetization, ${\mib n}$, \cite{com0}
as described by the Landau-Lifshitz-Gilbert equation, 
\begin{equation}
  \dot {\mib n} = \gamma_0 {\mib H}_{\rm eff} \times {\mib n} 
  + \alpha_0 \dot {\mib n} \times {\mib n} + {\mib t}_{\rm el}' , 
\label{eq:LLG}
\end{equation}
the following torques are known; 
\begin{equation}
 {\mib t}_{\rm el} '
= - ({\mib v}_{\rm s} \!\cdot\! {\mib \nabla})\,  {\mib n} 
  - \beta {\mib n} \times ({\mib v}_{\rm s} \cdot\! {\mib \nabla}) 
    \, {\mib n} 
  - \alpha \, ({\mib n} \times \dot {\mib n}) 
  - \frac{\delta S}{S} \, \dot {\mib n} . 
\label{eq:t_el'}
\end{equation}
 They include the celebrated spin-transfer torque 
(first term; ${\mib v}_{\rm s}$ being the \lq\lq spin-transfer velocity'') 
\cite{BJZ98,Ansermet04,Macdonald04,STK05,Xiao06}, 
the so-called \lq $\beta$-term' (second term) 
\cite{Zhang05,Thiaville05,Barnes05,TSBB06,KTS06,PT06,Duine07}, 
the Gilbert damping (third term), and 
the renormalization of spin (fourth term). \cite{com_dS}
 These torques are \lq\lq adiabatic'' in the sense that they are 
first order in space/time derivative \cite{com1,com2}. 
 In particular, the $\alpha$- and $\beta$-terms, which are classified 
as \lq\lq dissipative torques''\cite{Duine07} and 
arise from spin relaxation of conduction electrons \cite{Zhang05}, 
has been a current controversy 
\cite{TK04,Barnes05,Xiao06,TSBB06,KTS06,PT06,Duine07,Stiles07}. 
 This is due to the phenomenological nature of the argument, 
and thus their microscopic understanding and derivation is highly desired.

 There are several works on the microscopic determination of 
$\alpha$ and $\beta$ 
\cite{TSBB06,KTS06,PT06,Duine07,TFH04,Sakuma06,STBB06}. 
 For example, in ref.\cite{KTS06} (to be referred to as I), 
we have carried out a fully microscopic calculation 
based on the linear response of conduction electrons 
to small-amplitude magnetization fluctuations 
around a uniformly magnetized state. 
 However, application of the results to finite-amplitude dynamics 
requires an assumption of rotational symmetry on the spin torque, 
as exploited in writing the functional form of eq.(\ref{eq:t_el'}).

 In this paper, we present a formulation which is not restricted to 
the small-amplitude dynamics, and thus allows a 
\lq\lq first-principle'' derivation of the spin torque 
without any assumptions such as rotational symmetry.  
 This is achieved by working with an \lq\lq adiabatic'' spin frame, 
where an inhomogeneous (and time-dependent) magnetization 
is transformed to a homogeneous (and static) one with a price 
of the appearance of SU(2) gauge field. 
 This method based on gauge fields has been used in, {\it e.g.}, 
spin-fluctuation theory \cite{Korenman77}, 
quantum tunneling and resistivity of domain walls \cite{TF94&97}, 
and spin-transfer torque \cite{BJZ98,STK05}. 
 Extension of this method to dissipative 
torques ($\alpha$- and $\beta$-terms) is the aim of the present work.

 We study the same model as I, namely, the $s$-$d$ model 
consisting of conducting $s$ electrons and localized $d$ spins. 
 The total Lagrangian is given by 
$ L_{\rm tot} = L_S + L_{\rm el} - H_{sd}$, 
where $L_S$ is the Lagrangian for $d$ spins, and 
\begin{align}
  L_{\rm el} 
&= \int d{\mib r} \, c^\dagger \left[ 
     i \hbar \frac{\partial}{\partial t} 
   + \frac{\hbar^2}{2m} \nabla^2 
   + \varepsilon_{\rm F} - V_{\rm imp}  \right] c \, , 
\label{eq:L_el}
\\
  H_{sd} 
&= -M \int d{\mib r} \, {\mib n} (x) \!\cdot\! (c^\dagger {\mib \sigma} c)_x . 
\label{eq:H_sd}
\end{align}
 Here 
$c^\dagger = c^\dagger (x) 
 = (c^\dagger_\uparrow (x), c^\dagger_\downarrow (x))$ 
is the electron creation operator at $x = ({\mib r},t)$, and 
${\mib \sigma}$ is the vector of Pauli spin matrices. 
 We adopt a continuum description for $d$ spins, with a fixed 
magnitude $S$ and varying direction ${\mib n}$.

 The $s$ electrons are treated as a free electron gas in three 
dimensions subject to impurity potential 
$ V_{\rm imp} 
=  u \sum_i \delta ({\mib r} - {\mib R}_i) 
   + u_{\rm s} \sum_j {\mib S_j} \!\cdot\! {\mib \sigma} 
               \delta ({\mib r} - {\mib R}_j') $; 
the second term represents quenched magnetic impurities, 
which is aimed at introducing spin relaxation process. 
 The averaging over impurity spin direction is taken as 
$\overline{S_i^\alpha} = 0$ and 
\begin{equation}
  \overline{S_i^\alpha S_j^\beta} 
= \frac{1}{3} S_{\rm imp}^2 \delta_{ij} \delta^{\alpha\beta} .
\label{eq:SS}
\end{equation}
 Namely, we consider isotropic spin scatterings, 
or simply, \lq\lq isotropic impurities'', 
to focus on the essence of the gauge-field formulation. 
 More general case of \lq\lq anisotropic impurities''
[$\,${\it i.e.}, 
 $\overline{(S_i^x)^2} = \overline{(S_i^y)^2} \ne \overline{(S_i^z)^2}$ 
as considered in I$\,\,$] 
will be commented on at the end.

 To treat finite-amplitude dynamics of magnetization, we work 
with a local/instantaneous spin frame 
(called \lq\lq adiabatic frame'' below) 
for $s$ electrons whose spin quantization axis is taken 
to be the local/instantaneous $d$-spin direction, ${\mib n}$. 
 The electron spinor $a(x)$ in the new frame is related to the 
original spinor $c(x)$ as $c(x) = U(x) a(x)$, where $U$ is a 
$2\times 2$ unitary matrix satisfying 
$ c^\dagger ({\mib n} \!\cdot\! {\mib \sigma}) c 
 = a^\dagger \sigma^z a$. 
 We take $ U(x) = {\mib m} (x) \!\cdot\! {\mib \sigma}$ 
with 
${\mib m}=(\sin(\theta/2) \cos\phi, \sin(\theta/2) \sin\phi, 
           \cos(\theta/2))$ 
based on the parametrization 
${\mib n}=(\sin\theta \cos\phi, \sin\theta \sin\phi, \cos\theta)$.

 Since 
$\partial_\mu c = U^\dagger (\partial_\mu + U \partial_\mu U) a 
 \equiv U^\dagger (\partial_\mu +iA_\mu) a$, 
there arises from $L_{\rm el}$ a coupling 
${\cal H}_1 = \hbar \int d{\mib r} 
 [ {\mib A}_0 \cdot \tilde {\mib \sigma} 
 + {\mib A}_i \cdot \tilde {\mib j}_{{\rm s},i} ] 
 + {\cal O} ({\mib A}_i^2)$ 
of $a$-electrons with an SU(2) gauge field 
\begin{equation}
 A_\mu = -i U^\dagger (\partial_\mu U) = A^\alpha_\mu \sigma^\alpha 
\equiv {\mib A}_\mu \!\cdot {\mib \sigma} , 
\label{eq:A_UdU}
\end{equation}
in the new frame. 
 Here ${\mib A}_\mu \equiv {\mib m} \times \partial_\mu {\mib m}$ 
is a measure of temporal ($\mu=0$) or spatial ($\mu=1,2,3$) variation 
of magnetization, 
and couples to spin, 
$\tilde {\mib \sigma} (x) = (a^\dagger {\mib \sigma} a)_x $, 
and spin current, 
$\tilde {\mib j}_{{\rm s},i} 
= (\hbar /2mi) [a^\dagger {\mib \sigma} 
  \overset{\leftrightarrow}{\partial}_i a]$, 
in the new frame.

 The spin-torque density from $H_{sd}$ is given by 
\begin{equation}
 {\mib t}_{\rm el} (x) 
= M {\mib n} \times \langle {\mib \sigma} (x) \rangle_{\rm ne} 
= M {\cal R} \, (\hat z \times \langle \tilde {\mib \sigma} (x) 
  \rangle_{\rm ne}) ,
\label{eq:t_el}
\end{equation}
where ${\mib \sigma} (x) = (c^\dagger {\mib \sigma} c)_x$, and 
$\langle \cdots \rangle_{\rm ne}$ means averaging 
in nonequilibrium states with time-dependent ${\mib n}$ 
or steady current. 
 We have introduced a $3\times 3$ orthogonal matrix 
\begin{equation}
 {\cal R}^{\alpha \beta} = 2m^\alpha m^\beta - \delta^{\alpha \beta} ,
\label{eq:R_ab}
\end{equation}
representing the same rotation as $U$ but in a three-dimensional vector 
space. 
 Note that ${\cal R} \hat z = {\mib n}$, 
$c^\dagger {\mib \sigma} c
= {\cal R} \, (a^\dagger {\mib \sigma} a)$, and ${\cal R}^2=1$.

 Since the gauge field ${\mib A}_\mu$ contains a space/time derivative 
of magnetization, one may naturally formulate a gradient expansion 
in terms of ${\mib A}_\mu$ to calculate, {\it e.g.}, the 
torque (or spin polarization). 
 In particular, the adiabatic torques are obtained as the 
first-order terms in ${\mib A}_\mu$: 
\begin{equation}
 \langle \tilde {\mib \sigma}_\perp \rangle_{\rm ne} 
= \frac{2}{M} \left[ 
   a_\mu {\mib A}^\perp_\mu 
  + b_\mu (\hat z \times {\mib A}^\perp_\mu )  \right] .
\label{eq:sigma_perp_A}
\end{equation}
 Here 
$\tilde {\mib \sigma}_\perp 
 = \tilde {\mib \sigma} 
 - \hat z \, (\hat z \!\cdot\! \tilde {\mib \sigma})$ 
and 
$ {\mib A}^\perp_\mu
= {\mib A}_\mu - \hat z \, (\hat z \!\cdot\! {\mib A}_\mu) $ 
are the respective transverse components \cite{com3}, and 
the sums over $\mu = 0,1,2,3$ are understood.  
 From the identities, 
\begin{align}
& {\cal R} {\mib A}^\perp_\mu
= - \frac{1}{2} {\mib n} \times (\partial_\mu {\mib n}) , 
\ \ \ 
{\cal R} (\hat z \times {\mib A}^\perp_\mu) 
= \frac{1}{2} \partial_\mu {\mib n} , 
\label{eq:RA}
\end{align}
together with eq.(\ref{eq:t_el}), 
we see that eq.(\ref{eq:sigma_perp_A}) leads to the torque density 
$ {\mib t}_{\rm el} 
= a_0 \dot {\mib n} + ({\mib a} \!\cdot\! {\mib \nabla})\,  {\mib n} 
  + b_0 \, ({\mib n} \times \dot {\mib n}) 
  + {\mib n} \times ({\mib b} \cdot\! {\mib \nabla}) \, {\mib n} $, 
which is related to ${\mib t}_{\rm el}'$ of eq.(\ref{eq:t_el'}) via 
${\mib t}_{\rm el} = (\hbar S/a^3) \, {\mib t}_{\rm el}'$, 
with $a^3$ being the volume per $d$ spin.

 The coefficients $a_\mu$ and $b_\mu$ are expressed by 
response functions which are wavevector (${\mib q}$) 
and frequency ($\omega$) dependent, 
and the relation between 
$\langle \tilde {\mib \sigma}_\perp \rangle_{\rm ne}$ 
and ${\mib A}_\mu^\perp$ is nonlocal in general. 
 For adiabatic torques, however, it is sufficient 
to estimate them at ${\mib q} = {\mib 0}$ and $\omega = 0$, 
since the gauge fields already contain space/time derivative. 
 The torque then becomes a local function of ${\mib n}$ 
like eq.(\ref{eq:t_el'}). 

The following calculations are done with Green's functions, 
$G_{{\mib k} \sigma}(z) 
= (z - \hbar^2 {\mib k}^2/2m + \varepsilon_{{\rm F} \sigma} + i\gamma_\sigma 
 {\rm sgn} ({\rm Im} z))^{-1}$
where 
$ \gamma_\sigma  
 =  \hbar / (2\tau_\sigma) 
 =  \pi n_{\rm i}u^2  \nu_\sigma  
  + \pi n_{\rm s} u_{\rm s}^2 S_{\rm imp}^2
    ( 2 \nu_{\bar\sigma} + \nu_\sigma )/3 $ 
is the damping rate, 
and $\varepsilon_{{\rm F} \sigma} \equiv \varepsilon_{\rm F} + \sigma M$. 
 Here $n_{\rm i}$ ($n_{\rm s}$) is the concentration of normal 
(magnetic) impurities, and 
$\nu_\sigma = m \, k_{{\rm F} \sigma}/2\pi^2\hbar^2$ 
(with $k_{{\rm F} \sigma} = \sqrt{2m\varepsilon_{{\rm F} \sigma}}/\hbar$) 
is the density of states at 
$\varepsilon_{{\rm F} \sigma}$. 
 We calculate the torques 
in the lowest nontrivial order in 
$\gamma_\sigma / \varepsilon_{{\rm F} \sigma}$ and 
$\gamma_\sigma / M$, as in I \cite{com33}.

\begin{figure}[b]
  \begin{center}
  \includegraphics[scale=0.45]{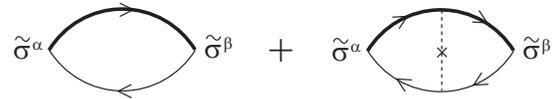}
  \vskip -2mm
  \end{center}
\caption{Transverse spin susceptibility, $\chi_\perp^{\alpha\beta}$, 
up to ${\cal O}(\gamma_\sigma )$, 
contributing to the spin torque related to the time 
variation of magnetization. 
 The thick (thin) solid line represents an electron line carrying 
Matsubara frequency 
$i\varepsilon_n + i\omega_\lambda$ ($i\varepsilon_n$).
 The dotted line with a cross represents scattering 
by static (quenched) impurities, either non-magnetic or magnetic. }
\label{fig:GF_a}
\end{figure}

{\bf Gilbert Damping:} 
 We first study the torques associated with the temporal 
variation of ${\mib n}$. 
 Assuming ${\mib n}$ to be uniform, ${\mib n} = {\mib n}(t)$, 
we focus on the $\mu = 0$ terms in eq.(\ref{eq:sigma_perp_A}).
 The coefficients $a_0$ and $b_0$ are related to 
the transverse spin susceptibility, $\chi_\perp^{\alpha\beta}$, as 
$\frac{2}{M}(-a_0 \delta^{\alpha\beta}_\perp 
 + b_0 \varepsilon^{\alpha\beta})
= \hbar \, \chi_\perp^{\alpha\beta} ({\mib q} = {\mib 0}, \omega = 0)$, 
whereas we obtain (Fig.1) \cite{KTS06} 
\begin{equation}
 \chi_\perp^{\alpha\beta} ({\mib 0}, 0) 
= \frac{\rho_{\rm s}}{M} \, \delta^{\alpha\beta}_\perp ,
\label{eq:chi_0}
\end{equation}
where $\rho_{\rm s} = n_\uparrow - n_\downarrow$ 
(with $n_\sigma = k_{{\rm F} \sigma}^3/6\pi^2$) 
is the $s$-electron spin density, 
$\delta^{\alpha\beta}_\perp \equiv \delta^{\alpha\beta} 
 - \delta^{\alpha z} \delta^{\beta z}$, 
$\varepsilon^{xy}=-\varepsilon^{yx}=1$, 
and 
$\varepsilon^{xx}=\varepsilon^{yy}=0$. 
 This leads to $\delta S = \rho_{\rm s} a^3/2 $ and $\alpha = 0$. 
 The Gilbert damping does not arise. 
 One may question the result (\ref{eq:chi_0}), but it seems robust 
since it assures the fact that, when the direction of ${\mib n}$ 
changes slightly, the equilibrium 
$s$-electron spin density 
follows it with the same magnitude $\rho_{\rm s}$.
 Thus the gauge-field method has an apparent difficulty with 
Gilbert damping.

 This difficulty can be resolved if we note that the impurity spins, 
which are static (quenched) in the original frame, 
become time-dependent in the adiabatic frame. 
 Namely, the spin part of $V_{\rm imp}$ is expressed as 
\begin{equation}
 {\mib S_j} \!\cdot\! c^\dagger {\mib \sigma} c
= \tilde {\mib S_j}(t)  \!\cdot\! a^\dagger {\mib \sigma} a 
\label{eq:Sc_SSa}
\end{equation}
where 
$\tilde {\mib S_j}(t) \equiv {\cal R}(t)^{-1} {\mib S_j} 
= {\cal R}(t) {\mib S_j}$, 
or 
\begin{equation}
 \tilde S_j^\beta (t') 
= {\cal R}^{\beta \beta'}(t') S_j^{\beta'} 
\label{eq:SS_RS}
\end{equation}
is the impurity spin in the adiabatic frame, which is time-dependent. 
 While the $s$-electron spin tends to follow the instantaneous $d$-spin 
direction ${\mib n}(t)$, it is at the same time 
pinned by the quenched impurity spins ${\mib S}_j$. 
 These two competing effects are represented by the time dependence 
of $\tilde {\mib S}_j (t)$ in the adiabatic frame. 
 The above observation (and difficulty) is based on the neglect 
of this time dependence.\cite{com4}

 To study the effect of time variation of 
$\tilde S_j^\alpha (t)$, we calculate 
$\langle \tilde \sigma_\perp^\alpha (t) \rangle$ 
perturbatively with respect to $\tilde S_j^\alpha (t)$. 
 The nontrivial contribution comes up at the second order 
(since $\overline{S_i^\alpha} = 0$). 
 Using the extension of the linear response theory to the second-order 
response \cite{Kubo}, we obtain 
\begin{align}
 \langle \tilde \sigma_\perp^\alpha (t) \rangle_{\rm ne} 
&= n_{\rm s} u_{\rm s}^2 
   \int_{-\infty}^\infty dt' \int_{-\infty}^\infty dt'' 
   \chi_{\alpha \beta \gamma}^{\rm R} (t-t',t'-t'') 
\nonumber \\
& \hskip 35mm \times 
   \overline{\tilde S^\beta (t') \tilde S^\gamma (t'')} , 
\label{eq:sigma_NLR}
\end{align}
where 
$ \chi_{\alpha \beta \gamma}^{\rm R} (t-t',t'-t'')
 = - (2\hbar^2)^{-1} \, \theta (t-t') \, \theta (t'-t'') \, 
   \langle [ [ \tilde \sigma_\perp^\alpha (t) , 
               \tilde \sigma^\beta (t') ] , 
               \tilde \sigma^\gamma (t'') ] \rangle 
 + \ (t' \leftrightarrow t'', \beta \leftrightarrow \gamma)$ 
is the retarded three-spin correlation function, here playing the 
role of nonlinear response function \cite{com5}.
 From eqs.(\ref{eq:SS}) and (\ref{eq:SS_RS}), we have
\begin{equation}
 \overline{\tilde S^\beta (t') \tilde S^\gamma (t'')}  
= \frac{1}{3} S_{\rm imp}^2 
   \left[ {\cal R}(t')  {\cal R} (t'') \right]^{\beta \gamma} .
\label{eq:S(t')S(t'')}
\end{equation}
 The correlation function 
$\chi_{\alpha \beta \gamma}^{\rm R} (t-t',t'-t'')$ 
will decay as a function of time separations, 
with a time scale of the spin relaxation time 
$\tau_{\rm s}$. 
 For $\omega \tau_{\rm s} \ll 1$, where $\omega$ is a typical 
frequency of magnetization dynamics, 
we may make a short-time approximation for the time-dependence 
of ${\mib n}$ (and ${\cal R}$), 
and retain the lowest nontrivial order in the expansion,
$ {\cal R}(t') = {\cal R}(t) + (t'-t) \dot {\cal R}(t) + \cdots$. 
 This leads to 
\begin{equation}
 {\cal R}(t')  {\cal R} (t'') 
= \hat 1 - (t'-t'') {\cal R}(t) \dot {\cal R}(t) + \cdots ,
\label{eq:R(t')R(t'')}
\end{equation}
where we have used 
$\dot {\cal R}(t) {\cal R}(t) 
 = - {\cal R}(t) \dot {\cal R}(t)$. 
 The identity 
\begin{equation}
 [ {\cal R}(t)  \dot {\cal R}(t) ] ^{\alpha \beta} 
= 2 \varepsilon^{\alpha \beta \gamma} A_0^\gamma 
\label{eq:RdR_A0}
\end{equation}
shows that we have found a contribution of the gauge field, 
which was not expected at the initial stage of the formulation 
when we introduced the gauge field, eq.(\ref{eq:A_UdU}).

 We rewrite eq.(\ref{eq:sigma_NLR}) in terms of Fourier components, 
\begin{equation}
 \langle \tilde \sigma_\perp^\alpha (\omega ) \rangle_{\rm ne} 
= n_{\rm s} u_{\rm s}^2 \int_{-\infty}^\infty 
    \frac{d\omega'}{2\pi} \, 
    \chi_{\alpha\beta\gamma}^{\rm R} (\omega, \omega') 
    \overline{\tilde S^\beta (\omega - \omega') 
              \tilde S^\gamma (\omega')} , 
\label{eq:sigma_NLR_omega}
\end{equation}
and expand $\chi_{\alpha\beta\gamma}^{\rm R} (\omega, \omega')$ 
with respect to $\omega$ and $\omega'$ as 
\begin{equation}
 \chi_{\alpha\beta\gamma}^{\rm R} (\omega, \omega') 
= A_{\alpha\beta\gamma} 
    - i\omega  B_{\alpha\beta\gamma} 
    - i\omega' C_{\alpha\beta\gamma} 
    + \cdots 
\label{eq:chi_abc_expansion}
\end{equation}
up to the first order, where 
$A_{\alpha\beta\gamma} 
 \equiv \chi_{\alpha\beta\gamma}^{\rm R} (0,0)$, 
and 
$B_{\alpha\beta\gamma}$ and $C_{\alpha\beta\gamma}$ 
are the coefficients. 
 Back to real time, we have 
\begin{align}
 \langle \hat \sigma_\perp^\alpha (t) \rangle_{\rm ne} 
& = n_{\rm s} u_{\rm s}^2 \biggl\{ 
   A_{\alpha\beta\gamma} 
     \overline{\tilde S^\beta (t) \tilde S^\gamma (t)} 
 + B_{\alpha\beta\gamma} \frac{d}{dt} 
     \overline{\tilde S^\beta (t) \tilde S^\gamma (t)} 
\nonumber \\
& \hskip 30mm 
 + C_{\alpha\beta\gamma}  
     \overline{\tilde S^\beta (t) \dot{\tilde S}^\gamma (t)} 
  + \cdots  \biggr\} 
\nonumber \\
& = \frac{1}{3} n_{\rm s} u_{\rm s}^2 S_{\rm imp}^2
    \left\{ 
    A_{\alpha\beta\beta} 
  + 2 C_{\alpha\beta\gamma} 
    \varepsilon^{\beta \gamma \lambda} A_0^\lambda (t) 
  + \cdots  \right\} . 
\label{eq:sigma_A0}
\end{align}
 Hence it suffices to calculate the $\omega'$-linear coefficient, 
$C_{\alpha\beta\gamma}$, of 
$\chi_{\alpha\beta\gamma}^{\rm R} (\omega, \omega')$.

\begin{figure}[b]
  \begin{center}
  \includegraphics[scale=0.45]{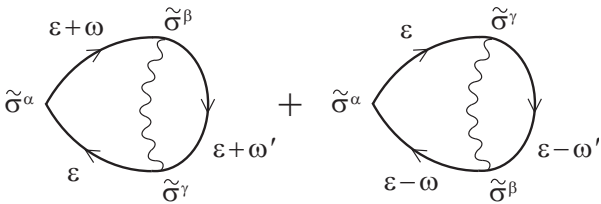}
  \vskip -2mm
  \end{center}
\caption{Contribution to the Gilbert damping in the lowest order 
in $\gamma_\sigma$. 
 The wavy line represents scattering from impurity spins, 
which are dynamically fluctuating in the adiabatic frame. 
 The running frequencies are indicated. }
\label{fig:GF_c}
\end{figure}

 The retarded function 
$\chi_{\alpha\beta\gamma}^{\rm R} (\omega, \omega')$ 
can be obtained from the Matsubara function 
\begin{align}
& \chi_{\alpha\beta\gamma} (i\omega_\lambda, i\omega_\lambda')
 = \frac{1}{2} \int_0^\beta d(\tau - \tau'') 
    \int_0^\beta d(\tau' - \tau'') \, 
\nonumber \\
& \hskip 5mm \times
   {\rm e}^{i\omega_\lambda  (\tau - \tau') 
                + i\omega_\lambda' (\tau' - \tau'')} \, 
    \langle {\rm T}_\tau [ 
        \tilde \sigma_\alpha (\tau) 
        \tilde \sigma_\beta (\tau') \tilde \sigma_\gamma (\tau'') ] 
    \rangle 
\label{eq:chi_abc_Matsubara}
\end{align}
by the analytic continuation, $i\omega_\lambda \to \omega + 2i\eta$ 
and $i\omega_\lambda' \to \omega' + i\eta$, 
where $\eta$ is a positive infinitesimal. 
 It is sufficient to consider the lowest-order process 
(with respect to the spin scattering), 
whose diagrammatic expression is shown for 
$\langle \tilde \sigma_\perp^\alpha (i\omega_\lambda) \rangle_{\rm ne}$ 
in Fig.\ref{fig:GF_c}. 
 The corresponding 
$\chi_{\alpha\beta\gamma} (i\omega_\lambda, i\omega_\lambda')$ 
is given by
\begin{align}
& \chi_{\alpha\beta\gamma} (i\omega_\lambda, i\omega_\lambda') 
 = \frac{1}{2} T\sum_n \sum_{{\mib k},{\mib k}'} 
  {\rm tr} \Bigl[ \sigma^\alpha 
   G_{\mib k}(i\varepsilon_n + i\omega_\lambda) 
   \sigma^\beta 
\nonumber \\
&  \times
   G_{{\mib k}'}(i\varepsilon_n+i\omega_\lambda') 
  \sigma^\gamma G_{\mib k}(i\varepsilon_n ) \Bigr] 
+ (i\omega_\lambda' \to i\omega_\lambda - i\omega_\lambda') ,
\label{eq:chi_abc_1}
\end{align}
where 
$G_{\mib k} = {\rm diag} (G_{{\mib k} \uparrow}, G_{{\mib k} \downarrow})$. 
 To leading order in $\gamma_\sigma$, we have 
\begin{equation}
 C_{\alpha\beta\gamma} 
= \frac{1}{4\pi} \sum_{{\mib k},{\mib k}'} 
 {\rm tr} \Bigl[ \sigma^\alpha G^{\rm R}_{\mib k} 
 \sigma^\beta \bigl( G^{\rm R}_{{\mib k}'} - G^{\rm A}_{{\mib k}'} \bigr)
 \sigma^\gamma G^{\rm A}_{\mib k} \Bigr] 
- (\beta \leftrightarrow \gamma) ,
\label{eq:C_abc}
\end{equation}
and thus 
$ C_{\alpha\beta\gamma} = - (\pi/2M)  \nu_+^2 
 (\delta^{\alpha\beta}_\perp \delta^{\gamma z} 
 - \delta^{\alpha\gamma}_\perp \delta^{\beta z})$, 
where $\nu_+ = \nu_\uparrow + \nu_\downarrow$. 
 Therefore, we obtain 
\begin{equation}
 \langle \tilde {\mib \sigma}_\perp (t) \rangle_{\rm ne} 
= - \frac{2\pi}{3M} n_{\rm s} u_{\rm s}^2 S_{\rm imp}^2 
  \nu_+^2 \, (\hat z \times {\mib A}_0^\perp (t))  , 
\label{eq:sigma_iso_result_vec}
\end{equation}
leading to the Gilbert damping with 
\begin{equation}
 \alpha 
 = \frac{\pi}{3} n_{\rm s} u_{\rm s}^2 S_{\rm imp}^2 \nu_+^2 
  \times \frac{a^3}{S} , 
\label{eq:alpha_iso_result}
\end{equation}
in agreement with the result of I for isotropic impurities.

 The above calculation provides us a new picture of Gilbert damping. 
 In the adiabatic frame, impurity spins are dynamically 
fluctuating and disturb the $s$-electron spins. 
 Because of this disturbance, $s$ electrons admit a low-energy spectral weight 
in the transverse spin channel (at wavevector ${\mib q} = {\mib 0}$), 
which causes Gilbert damping.

\begin{figure}[b]
  \begin{center}
  \includegraphics[scale=0.33]{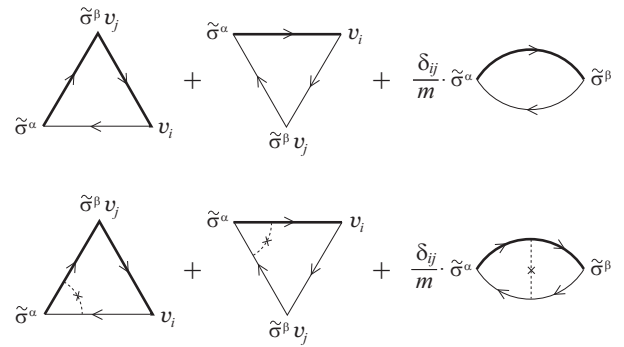}
  \vskip -2mm
  \end{center}
\caption{
 The $E_i$- and $A^\beta_j$-linear coefficient, 
$\tilde K_{ij}^{\alpha\beta}$, of the transverse spin polarization, 
$\langle \tilde\sigma^\alpha_\perp ({\mib q}) \rangle_{\rm ne}$, 
in the presence of current flow ($E_i$) and magnetization texture 
($A^\beta_j$). 
 The vertices with $v_i$ and $\tilde \sigma^\beta v_j$ 
are associated with $E_i$ and $A^\beta_j$, respectively. 
 Other graphical meanings are the same as Fig.1.}
\label{fig:GF_b}
\end{figure}

{\bf Current-Induced Torques:}
 We next calculate the spin torques induced by 
electric/spin currents in the presence of spatial variation of 
${\mib n} = {\mib n}({\mib r})$. 
 These torques correspond to terms in 
$\langle \tilde {\mib \sigma}_\perp \rangle_{\rm ne}$ 
proportional to the spatial components of the gauge field, 
${\mib A}_i^\perp$ 
(and $\hat z \times {\mib A}_i^\perp$). 
 The texture ${\mib n}$ (and hence ${\mib A}_i$) 
is assumed to be static here.

 We produce a steady current state by applying a d.c. electric 
field ${\mib E}$, and first calculate the linear response of 
$\tilde \sigma_\perp^\alpha$ to ${\mib E}$. 
 Using Kubo formula, \cite{Kubo} one can derive 
\begin{align}
& \langle \tilde\sigma^\alpha_\perp ({\mib q}) \rangle_{\rm ne} 
  =  \lim_{\omega \to 0}
    \frac{ \tilde K_i^\alpha ({\mib q},\omega +i0) 
         - \tilde K_i^\alpha ({\mib q},0)}{i\omega} \,  E_i , 
\\
& \ \ \ \ 
 \tilde K_i^\alpha ({\mib q}, i\omega_\lambda ) 
 = \int_0^\beta d\tau \, {\rm e}^{i\omega_\lambda \tau} \, 
    \langle \, {\rm T}_\tau \, \tilde\sigma_\perp^\alpha ({\mib q}, \tau) 
            \, J_i \, \rangle . \ \ \ 
\label{eq:Ka_Kubo_1}
\end{align}
 Here 
$J_i = - e \sum_{{\mib k}} v_i  
   \, c_{{\mib k}}^\dagger 
      c_{{\mib k}}^{\phantom{\dagger}} 
= - e \sum_{{\mib k}} v_i
   \, a_{{\mib k}}^\dagger 
      a_{{\mib k}}^{\phantom{\dagger}} 
  - (e\hbar/m) \sum_{{\mib k},{\mib q}'} A^\beta_{{\mib q}',i} 
   a_{{\mib k} + {\mib q}'}^\dagger \sigma^\beta 
   a_{{\mib k}}^{\phantom{\dagger}} $, 
with $v_i= \hbar k_i/m$, 
is the total electric (or charge) current, and  
the average $\langle \cdots \rangle$ is taken in the thermal equilibrium 
state determined by $L_{\rm el} - H_{sd}$. 
 We extract the gauge field in the first order, and write as 
\begin{equation}
 \tilde K_i^\alpha ({\mib q}, i\omega_\lambda ) 
 = e\hbar \, \tilde K_{ij}^{\alpha\beta} (i\omega_\lambda ) 
   A^\beta_{{\mib q},j} . 
\label{eq:Ka_Kab_Ab_2}
\end{equation}
 Diagrammatic expressions for $\tilde K_{ij}^{\alpha\beta}$ are 
shown in Fig.\ref{fig:GF_b}. 
 The calculation is rather straightforward, yielding 
\begin{equation}
  {\mib a} 
= \frac{\hbar}{2e} {\mib j}_{\rm s} , 
\ \ \ 
  {\mib b} 
=  \frac{2\pi}{3M} \!\cdot\! 
     n_{\rm s}u_{\rm s}^2 S_{\rm imp}^2 \nu_+ \!\cdot\!
     \frac{\hbar}{2e} {\mib j}_{\rm s} . 
\label{eq:ab}
\end{equation}
 Here 
${\mib j}_{\rm s} = \sigma_{\rm s}{\mib E}
 = {\mib j}_\uparrow - {\mib j}_\downarrow $ 
is the spin current, 
with $ \sigma_{\rm s} = (e^2/m) 
(n_\uparrow \tau_\uparrow - n_\downarrow \tau_\downarrow)$ 
being the \lq\lq spin conductivity".\cite{com9} 
 The result (\ref{eq:ab}) also agrees with I for isotropic 
impurities.

 Before closing, let us comment on the case of anisotropic impurities. 
 For the Gilbert damping, we can reproduce the result of I 
if we take the anisotropy axis to be ${\mib n}$.\cite{com8} 
 As for the $\beta$-term, we could only reproduce the 
first term of the previous result [eq.(25) or (26) of I], 
and not the second term.  
 There seems to be a key contribution to the torque from $V_{\rm imp}$, 
but we do not succeed in reproducing the result of I at present. 
 While this puzzle needs to be clarified, we would like to stress that 
the essence of the gauge-field method is 
already contained in the isotropic case as presented in this letter.

In summary, we have developed a theoretical scheme for the adiabatic 
spin torques in terms of gauge fields associated with the adiabatic 
spin frame of conduction electrons. 
 We have found an unexpected contribution of the gauge field 
from the source terms of spin relaxation, 
which is essential to give the Gilbert damping. 
 The message here is that the gauge-field method does work also for 
dissipative torques ($\alpha$- and $\beta$-terms), 
if we are careful enough and faithful to the microscopic model. 
 This method will be readily applicable to other type of 
spin-relaxation mechanisms such as the ones originating from 
various type of spin-orbit coupling.

\section*{Acknowledgment}

We would like to express our deep 
gratitude to Gen Tatara for his invaluable and fruitful interaction 
throughout this work. 
 We also would like to thank H. Fukuyama for his kind advice. 
 This work is supported by a Grant-in-Aid for Scientific Research 
(No. 16540315) from the Japan Society for the Promotion of Science.


\begin{thebibliography}{99}
\bibitem{com0}
 We define ${\mib n}$ to be a unit vector in the direction of spin, 
which is opposite to the direction of magnetization.
\bibitem{com_dS}
 This term renormalizes the magnitude of spin from the localized 
component ($S$) to $S_{\rm tot} = S + \delta S$, where $\delta S$ 
is the contribution from $s$ electrons. 
 One can eliminate this (fourth) term by multiplying a factor 
$S/S_{\rm tot}$ to all terms in eq.(\ref{eq:t_el'}).
\bibitem{com2}
 There is also a non-adiabatic torque, ${\mib \tau}_{\rm na}$, 
due to electron reflection, which is spatially nonlocal and 
oscillating. 
 We do not study this torque in this letter; please 
refer to refs.\cite{Xiao06,WV07,TTKSNF06}
\bibitem{com3}
 We use the vector (bold italic) notation for the spin component. 
 The space-time components are indicated by subscripts such as 
$\mu, \nu$ $(=0,1,2,3)$ and $i,j$ $(=1,2,3)$.
\bibitem{com4}
 Neglecting the time dependence of $\tilde S^\alpha_i (t)$ 
in the calculation of the coefficients of ${\mib A}_0^\perp$ 
will be legitimate for adiabatic torques, 
since ${\mib A}_0^\perp$ already contains a time derivative.
\bibitem{com5}
  To avoid complexity of the notation, and to focus on the essence of 
the story on Gilbert damping, we have dropped the spatial coordinates 
in the definition of $\chi_{\alpha\beta\gamma}$; 
$\tilde \sigma^\beta$ and $\tilde \sigma^\gamma$ are evaluated 
at the impurity position ${\mib R}_j$, which is to be averaged over, 
whereas $\tilde \sigma_\perp^\alpha$ is at the same point ${\mib r}$ 
as ${\mib t}_{\rm el}$ is evaluated at.
\end{thebibliography}
\end{document}